\documentstyle[prl,aps,preprint]{revtex}
\begin {document}
\draft
\title{An elementary mode coupling theory of random heteropolymer dynamics\\
\small{(glass transition / protein folding / Langevin equation )}\\
\small{Classification: Physics, Biophysics} 
      }
\author{Shoji Takada, John J. Portman, and Peter G. Wolynes}
\address{School of Chemical Sciences and Department of Physics, University of Illinois, Urbana, IL, 61801\\
{\it Contributed by Peter G. Wolynes} \\ 
Abbreviation: MCT, mode coupling theory ; RHP, random heteropolymer ; \\
DFI, dynamic functional integral.
}
\date{\today}
\maketitle
 

\begin{abstract}
{The Langevin dynamics of a random heteropolymer and its dynamic glass transition 
are studied using elementary mode coupling theory.
Contrary to recent reports using a similar framework, 
a discontinuous ergodic-nonergodic phase
transition is predicted for all Rouse modes at a finite temperature $T_{\rm A}$.
For sufficiently long chains, $T_{\rm A}$ is almost independent of chain 
length and is in good agreement with the value previously estimated by 
a static replica theory.}
\end{abstract}


\noindent
The self-organization of evolved biopolymers such as foldable proteins 
ultimately depends upon the chain dynamics of heteropolymers.
The modern statistical mechanical view using energy landscape ideas 
focuses on the analogy of folding with phase transitions in finite 
size systems and exploits to a large extent our understanding of the 
quasi-thermodynamic and static features of both regular systems 
and frustrated disordered systems such as spin glasses\cite{Bryngelson95}. 
The connection with detailed analytical theories of the time dependence 
of fluctuational motions of a randomly interacting chain has only recently 
received attention\cite{Kinzelbach,Timoshenko,Roan,Thirumalai}. 

For completely random heteropolymers (RHP) the quasi-thermodynamic analyses
\cite{Takada96,Takada97}
using replica techniques give results parallel to those for other systems 
with random first order phase transitions such as the mean field Potts 
spin glasses\cite{Kirkpatrick87}. 
An important feature of theories of random first order transitions is 
the presence of two transitions; one is static, while the other is dynamic 
and generally occurs at a higher temperature. The dynamical
 transition signals a crossover to a motional mechanism 
involving {\em a}ctivated motions\cite{Kirkpatrick87,Kirkpatrick} 
(We denote this 
dynamical transition temperature $T_{\rm A}$).
For RHP's two different kinds of approximate 
quasi-static theories based on replicas\cite{Takada96,Takada97} and on the 
generalized random energy model\cite{Plotkin} that takes into account 
correlation in the energy landscape yield these two transitions. 
Replica based techniques also yield estimates for the free energy 
barrier heights of activated motions between the two 
characteristic temperatures
\cite{Takada96,Takada97}.
Purely dynamical theories based on mode coupling ideas generally yield 
a transition in harmony with the quasi-static analysis of 
the dynamical transition which is in some sense a spinodal. 
While mode coupling theory (MCT) is perturbatively well defined for 
spin systems with long range random interactions\cite{Fischer,Kirkpatrick}, 
there are various versions 
developed in the theory of fluids\cite{Kawasaki} and polymers\cite{Schweizer} that 
are forced to make uncontrolled approximations. 
On the basis of two such calculations, the validity of the emerging 
picture of the dynamics and the energy landscape of RHP's
have been questioned\cite{Roan}.

Roan and Shakhnovich derive a mode coupling theory of the RHP
and explicitly solve it for a polymer in a good solvent\cite{Roan}. 
It is no surprise that an uncollapsed chain has no dynamical transition, 
but the authors further claim that this is actually a structural feature of the 
RHP mode coupling equations independent of the state of collapse.
Thirumalai et al.\cite{Thirumalai} derive another 
set of mode coupling equations 
for a somewhat different model that exhibits no static replica symmetry breaking\cite{Garel}
and conclude there is a dynamical transition with a transition temperature that
depends on the length scale of the motional mode considered\cite{Thirumalai}.
A numerical treatment of self-consistent equations for heteropolymer collapse, 
on the other hand, does show evidence for a dynamic freezing transition
\cite{Timoshenko}.
The inconsistency of these results with each other and 
in the first two cases with replica calculations is disturbing. 
The technical intricacy of these mode coupling calculations is a barrier to 
understanding their inconsistencies. We have therefore derived mode 
coupling equations using `elementary' methods like those used decades ago 
in the theory of critical phenomena\cite{Kawasaki}. The resulting equations differ 
in some respects from those of the earlier workers but clearly yield 
transitions in harmony with the quasi-static results.
These equations also lead to an understanding of how the dynamical freezing 
depends on chain length and state of collapse as well as 
how the dynamics varies 
for different modes of chain motion.

We consider a standard Hamiltonian for $N$ interacting beads
\begin{equation}
H=k_B T\sum_{i=1}^N ({\bf r}_{i+1}-{\bf r}_i)^2 +1/2\sum_{i\ne j} b_{ij}u(\Delta {\bf r}_{ij})
+V_{\rm ex}-\sum {\bf h}_i(t)\cdot{\bf r}_i(t),
\end{equation}
where ${\bf r}_i$ are the bead locations, 
$b_{ij}$ is chosen as Gaussian random with mean 
$b_0$ and variance $b^2$, $\Delta {\bf r}_{ij}=
{\bf r}_i-{\bf r}_j$, and $u({\bf r})$ is the two-body 
interaction chosen as a Gaussian $\exp -{\bf r}^2/\sigma^2$.  
$V_{\rm ex}$ is the excluded volume term which usually contains three body 
terms, but it is enough, at the level of the present analysis for phantom chain, 
to replace it with an effective Gaussian confinement 
term $V_{\rm conf}=k_BTB\sum {\bf r}_i^2$, with the constant $B$ chosen so that the radius 
of gyration $R_g$ becomes the physically required value determined by the packing fraction,
$\eta$\cite{onB}.
${\bf h}_i$ is the external force introduced for convenience in the 
derivation of the response function.
For technical simplicity, we adopt a ring polymer model where ${\bf r}_{N+1}={\bf r}_1$.
The Langevin equation for the beads is 
\begin{equation}\label{eq:LE}
\Gamma^{-1}\partial_t{\bf r}_i =-\partial \beta H /\partial{\bf r}_i 
+{\bf \xi}_i
\end{equation}
Here 
${\bf \xi}_i(t)$ is a Gaussian random force 
for which the first two moments are given by
 $\langle {\bf \xi}_i(t)\rangle =0$ and 
$\langle {\bf \xi}_i(t) {\bf \xi}_j(t^\prime)\rangle=2\Gamma^{-1}
\delta_{ij}\delta(t-t^\prime) {\bf 1}$ ({\bf 1} is the 3 by 3 unit matrix),
$\Gamma$ determines the microscopic time scale
(it will be set to unity) and $\beta=1/k_B T$ as usual.
Since our main interest is in collapsed states where hydrodynamic effects are less 
important, we ignore them for simplicity although their inclusion is not very hard.
A more essential simplification of 
our treatment is the neglect of 
entanglement effects. An uncrossable chain will possess more friction
than the phantom chain treated here\cite{Schweizer}, modifying the dynamical transition 
temperature.

A standard tool for dealing with the average over the quenched randomness in 
the Langevin dynamics has been the dynamic functional integral (DFI) 
formalism\cite{Fischer,Kirkpatrick}. Although this approach is generally
accepted for the infinite range spin model, direct application of the
formalism to RHP models is questionable for two
reasons:
1)The heteropolymer interaction is short range so that the mean 
field approximation is not exact; in such a situation, there exists some ambiguity 
with the DFI formalism in its use of steepest descents. 
2)In contrast to spin models, the Jacobian appearing in the DFI 
formalism of the heteropolymer model depends on the randomness. 
This point was neglected in previous treatments.
The usual formal identity for averaging over randomness 
in the Hamiltonian for a spin-glass model is no longer precisely 
true. Thus, we take an alternative and simpler route here; we use a
perturbation theory in the randomness $b_{ij}$ along  
with a self-consistent prescription that corresponds 
to a particular resummation of higher order terms. 
In the case of  the infinite range $p$-spin model 
(including the Sherrington-Kirkpatrick model), this procedure
still leads to exactly the same results as the DFI formalism\cite{Kirkpatrick}.

We sketch the perturbation scheme briefly since details 
will be published elsewhere\cite{Portman}. 
The Langevin equation, Eq (\ref{eq:LE}), can be expressed as
\begin{equation}\label{eq:PLE}
{\bf r}_{l}(t) = \sum_i G_{0,li}*({\bf \xi}_i + {\bf h_i}
-\beta/2{\partial\over\partial {\bf r}_i}
\sum_{kj}b_{kj} u\left( \Delta {\bf r}_{kj}) \right),
\end{equation}
where $*$ represents the time
convolution and $G_{0,ij}(t-t^\prime)$ is the 
$0$-th order response function $\partial \langle {\bf r}_i^{(0)}(t)\rangle/\partial {\bf h}_j(t^\prime)$.
Inserting the $0$-th order solution ${\bf r}_l^{(0)}=\sum G_{0,li}* {\bf \xi}_i$ 
into the right hand side of eq.(\ref{eq:PLE}) gives the first order solution 
${\bf r}_l^{(1)}= {\bf r}_l^{(0)}-\beta/ 2 \sum_i G_{0,li}*{\partial /\partial{\bf r}^{(0)}_i}
\sum_{kj} b_{kj} u\left(\Delta{\bf r}^{(0)}_{kj}\right)$.
We insert this once again into eq.(\ref{eq:PLE}), yielding the second order solution.
After taking an average over $b_{ij}$, we multiply the resulting equation
by $G_0^{-1}$ which yields 
\begin{equation}\label{eq:r2}
\Gamma^{-1}\partial_t{\bf r}^{(2)}_i=-\sum K_{ij}
{\bf r}^{(2)}_j(t) 
+\mu \sum_j \int_{-\infty}^{t}{\rm d} t^\prime 
\Delta G_{0,ij}(t-t^\prime)\nabla\nabla u\left(\Delta{\bf r}^{(0)}_{ij}(t)\right)\cdot
\nabla u\left(\Delta{\bf r}^{(0)}_{ij}(t^\prime)\right)
+{\bf \xi}_i(t),
\end{equation}
where $K_{ij}$ is the harmonic constant matrix including the confinement term,
\newline$K_{ij}{\bf 1}\equiv\partial^2/\partial {\bf r}_i\partial {\bf r}_j 
k_B T\left[ \sum ({\bf r}_{i+1}-{\bf r}_i)^2 + B\sum {\bf r}_i^2\right]$,
$\mu=(\beta b)^2$, and $\Delta G_{0,ij}=G_{0,ii}-G_{0,ij}-G_{0,ji}+G_{0,jj}$.
The random noise correlation can also be calculated by perturbation theory. 
With the use of the relation $C_{lm}=\sum_{ij}G_{li} * \langle {\bf \phi}_i{\bf \phi}_j \rangle * G_{mj}$, 
we can derive the expression for the correlations up to 
second order, $C^{(2)}$, which includes 
$\langle {\bf \phi}_i(t){\bf \phi}_j (t^\prime)\rangle^{(2)}$.

A part of higher order terms can be taken into account by 
employing a self-consistent prescription.
To this end, we first expand the memory kernel of eq.(\ref{eq:r2}) 
in $\Delta {\bf r}_{ij}(t^\prime)$. Then,
in the spirit of Kawasaki's derivation of MCT 
of critical dynamics, we replace ${\bf r}_i^{(0)}$ in the 
right hand side by ${\bf r}_i$ and $G_0$ by the 
perturbed response function, $G$, giving a renormalized 
Langevin equation,
\begin{equation}\label{eq:RGLE}
\Gamma^{-1}{\rm d}{\bf r}_i/{\rm d}t=-\sum K_{ij}
{\bf r}_j(t) 
+\mu \sum_j \int_{-\infty}^{t}{\rm d} t^\prime 
\Delta G_{ij}(t-t^\prime){\cal C}_{ij}( t-t^\prime)
\Delta {\bf r}_{ij}(t^\prime)+{\bf \phi}_i(t),
\end{equation}
where
${\cal C}_{ij}=\langle \partial \left[ \nabla\nabla u\left( \Delta{\bf r}_{ij}(t)\right)\cdot 
\nabla u\left(\Delta{\bf r}_{ij}(t^\prime)\right)\right]/ 
\partial \Delta {\bf r}_{ij}(t^\prime)\rangle$. 
In deriving this equation we used the isotropic symmetry of the model.
The random force ${\bf \phi}_i(t)$  is also renormalized in the same way. 
In the expression for $\langle {\bf \phi}_i(t){\bf \phi}_j (t^\prime)\rangle^{(2)}$, 
replacement of ${\bf r}_i^{(0)}$ by ${\bf r}_i$ leads to the
colored noise correlation function, 
\begin{equation}\label{eq:RGnoise}
\langle {\bf \phi}_i(t) {\bf \phi}_j(t^\prime) \rangle
=2\Gamma^{-1}{\bf 1}\delta_{ij}\delta(t-t^\prime)
+\mu\left[ \delta_{ij}\sum_k M_{ik}(t-t^\prime)-
M_{ij}(t-t^\prime)\right],
\end{equation}
where 
$M_{ij}=\langle \nabla u\left( \Delta{\bf r}_{ij}(t)\right) 
\nabla u\left(\Delta{\bf r}_{ij}(t^\prime)\right)  \rangle$ is the force-force correlation
on different beads. 
In calculating ${\cal C}$ and $M$ we approximate 
the stochastic process as Gaussian. The explicit expression for $M$ becomes 
$M_{ij}(t)=\left[(1+s\Delta C_{ij}(0))^2-(s \Delta C_{ij}
(t))^2\right]^{-5/2}\Delta C_{ij}(t)$, where 
$\Delta C_{ij}=C_{ii}-C_{ij}-C_{ji}+C_{jj}$ and $s=2/\sigma^2$.
For the ergodic phase,  with the time-translational symmetry and 
the relation 
$\left[\partial_t \Delta C_{ij}(t-t^\prime)\right]{\cal C}_{ij}(t-t^\prime)=\partial_tM_{ij}(t-t^\prime)$, 
we can verify that the fluctuation-dissipation theorem indeed holds. 

From Eqs (\ref{eq:RGLE}) and (\ref{eq:RGnoise}), 
a straightforward generalization of 
ref.\cite{Kirkpatrick}'s scheme for the $p$-spin model leads to 
a closed set of integro-differential equations for the correlation 
functions. In the present case, 
due to the sequence-translational symmetry of ring polymers, it is 
more transparent to write the equation in the Rouse mode representation. 
Namely, the Fourier transform of the correlation function with respect to the 
sequence satisfies the equation
\begin{equation}\label{eq:tprop}
\Gamma^{-1}\partial_t c_p(t)+c_p(t)/c_p(0)+\mu \int_0^t dt^\prime \partial_t c_p(t-t^\prime) 
\left[  m_0(t^\prime)-m_p(t^\prime)  \right]=0,
\end{equation}
where $c_p(t)={\rm FT}_{(i-j)\rightarrow p}[C_{ij}(t)]$ is the Rouse mode
correlation function and
$m_p(t)= {\rm FT}_{(i-j)\rightarrow p}\left[M_{ij}(t)\right]$ is a 
nonlinear function of all of the correlations $\{ c_{p^\prime}\}$. 
The equal time correlation function $c_p(0)$ 
obeys 
\begin{equation}\label{eq:scc0}
c_p(0)^{-1}={\rm FT}_{(i-j)\rightarrow p}[K_{ij}]-\mu[m_0(0)-m_p(0)]\equiv \kappa_p. 
\end{equation}

If we introduce the normalized Edwards-Anderson (EA) Rouse-Zimm order parameters 
$\tilde{q}_p=\lim_{t\rightarrow\infty}c_p(t)/c_p(0)$, 
 the self-consistent equation for $\tilde{q}_p$ becomes
\begin{equation}\label{eq:scea}
\tilde{q}_p / (1-\tilde{q}_p)=(\mu  / \kappa_p) \left[m_0(\infty)-m_p(\infty) \right]\equiv F_p,
\end{equation}
which is isomorphous to the equations for the MCT of structural glasses\cite{Gotze}.
Notice that if $F_p\ne 0$ the RZ modes have a static offset indicating
the trapping in a local minimum 
and that all $p$ modes are coupled in evaluating
$m_p(\infty)$. Thus, clearly, the dynamic glass transition, 
if any, should occur
simultaneously for all modes in this analysis.
In general, the dynamic glass transition can be obtained by two 
steps.  We first solve eq.(\ref{eq:scc0}) for $c_p(0)$ and then 
look for a non-trivial solution of eq.(\ref{eq:scea}).
Assuming $c_p(0)$, a purely static quantity, does not exhibit 
any singular behavior\cite{Kirkpatrick,Gotze}, 
 we use an unperturbed confined Rouse value 
(i.e.\, ${\rm FT}_{(i-j)\rightarrow p}[K_{ij}]$) for $c_p(0)$ here 
avoiding the first step. 

Fig.1 shows EA parameters $\tilde{q}_p$ as a function of $\mu$ with 
$p=1, 50, 200$ and $512$ for a $1024$-mer with parameters $\sigma=1$ and $\eta=0.8$.
The EA parameters indeed exhibit a discontinuous (first order) transition at 
a critical value denoted by $\mu_{\rm A}$ for all $p$ modes.
Increasing the chain length up to 32768, we numerically show that 
$T_{\rm A}$ defined by $\mu_{\rm A}=(b/k_BT_{\rm  A})^2$ converges to a finite 
value $T_{\rm A}=0.3b$(see Fig.2), which is in good agreement with the value $0.292b$ 
estimated by the static replica theory for essentially the same model\cite{Takada97}.
This value depends little on the choice of $\eta$ and thus the agreement 
with the replica result is quite robust; 
with $\eta=0.6$ and $\eta=1.0$ we found 
$T_{\rm A}=0.28b$ and $T_{\rm A}=0.32b$, respectively. 

Scanning a wide range of chain lengths,$N$, the self-consistent equations 
yield two different transitions depending on chain length and state of collapase (Fig.2).
For collapsed phase (e.g., $\eta=0.8$), with short chain length, 
the EA parameters first achieve a small 
nonzero value (of order $10^{-2}$) giving a weakly nonergodic phase at a 
parameter value $\mu_c$ (dashed curve in Fig.2). 
A stronger nonergodic phase, 
in which EA parameters become of order unity ($\mu_{\rm A}$), 
appears for collapsed state 
with chain length longer than $128$ (solid curve in Fig.2). 
$\mu_c$ and $\mu_{\rm A}$ coalesce at $N\sim 512$ above which no 
weakly nonergodic phase exists. 
Under $\Theta$ solvent conditions (i.e., $\eta=0$), 
on the other hand, we do not find any second transition and 
$\mu_c$ continues to increase with $N$.
Since the statistical dynamical theory presented here is inherently appropriate for sufficiently large 
systems, we interpret the transition at $\mu_{\rm A}$ to correspond 
to the dynamical transition 
found by the replica approach\cite{Takada96,Takada97}. 
The weak transition ($\mu_c$) found here only for short chains is 
fragile and could be an artifact of the model 
and  we postpone its detailed interpretation.

Fig.3 plots $q_p=\lim_{t\rightarrow\infty}c_p(t )$ as a function of $p$ as well as 
its inverse Fourier transformed frozen static displacement $Q_{i-j}$. 
Since the $p=0$ mode 
is purely diffusive, we drop this component. 
Roughly, the frozen fluctuations $q_p$ and $Q_{i-j}$'s vary 
with $p$ and $i-j$ in a manner quite similar to the equal time fluctuations 
$c_p(0)$ and $C_{i-j}$, respectively. Although the frozen modes are 
quite localized in the bead 
representation, the long wavelength fluctuations corresponding to small $p$ 
are considerable.

Which mode is dominantly responsible for the transition?
The reduction theorem of G\"{o}tze\cite{Gotze} suggests that 
within MCT only one 
particular mode causes the instability of the nonergodic glass phase and 
the eigenvector of the stability matrix 
$S_{pp^\prime}=\partial F_p/\partial \tilde{q}_{p^\prime}$ 
with the largest eigenvalue corresponds to the most {\em dangerous} mode.
Dashed curves in Fig.3 show the right-eigenvector of this dangerous mode in 
the Rouse and bead representation, $\nu_p$ and $\nu_{i-j}$, 
have similar behavior
to $c_p(0)$ and $C_{i-j}$, respectively.

Eq.(\ref{eq:tprop}) can directly be integrated to get the explicit time dependence of $c_p(t)$. 
As in the MCT for structural glasses, critical slowing down is 
expected to be found as $T_{\rm A}$ is approached.
Details will be given elsewhere\cite{Portman}.

As mentioned above, the analysis of Roan and Shakhnovich yielded 
no singular behavior in the relaxation spectrum. 
There are several points in their analysis which differ from the present one 
but primarily their analysis is a lowest order perturbation with 
respect to $b_{ij}$. However, it is clear from our argument that 
the self-consistent nature of the MC calculation is essential for 
the dynamic glass transition 
and obviously this is included at a higher order in perturbation.
This is a principal reason for their discordant result. 
Thirumalai et al.'s approach is quite similar in spirit to ours,
but they analyze an uncoupled mode equation analogous to eq.(\ref{eq:tprop})
which yields a scale-dependent transition temperature.  
We believe that it is more accurate to include the coupling of 
different Rouse modes, forcing them to freeze simulutaneously.
The consistency with the static replica calculation buttresses the belief.

We emphasize again that 
the dynamic glass transition studied here by elementary MCT is, in reality, 
a crossover being
smeared out due to entropic droplets as discussed in refs.\cite{Kirkpatrick87,Takada97}.
Thus, $T_{\rm A}$ should not be viewed as a crisp phase transition, but 
a characteristic temperature at which the nature of chain dynamics changes 
qualitatively from renormalized free chain dynamics to activated escape 
from traps. 
Above $T_{\rm A}$, 
the so-called {\em cage effect} due to the 
mode coupling addressed here renormalizes the Rouse relaxation through internal 
friction yielding slow dynamics 
analogous to the $\alpha$ relaxation of supercooled liquids,
while below $T_{\rm A}$(but still above the static glass transition temperature) 
escape from localized free energy minima 
by (local) thermal activations, i.e.\ entropic droplets, controls the dynamics.

In applying these results to protein folding we need to consider that proteins 
are not entirely random but have evolved to satisfy the 
{\em minimal frustration principle}\cite{Bryngelson95}
so as to avoid being trapped in nonnative local minima. 
The sequence is designed so that energetically favorable but structurally 
incorrect amino acid contacts are 
minimized.
Thus, effects of the random heteropolymer's interaction considered here are 
superimposed on the flow of configurations through a global {\em funnel}
leading to the native structure\cite{Bryngelson95}. 
Qualitatively, the folding transition temperature $T_{\rm F}$ should be 
larger than the static glass transition $T_{\rm K}$ for fast folding and 
may perhaps be even larger than $T_{\rm A}$. If this is the case, 
folding can be viewed as a diffusion process in an order parameter space, 
where slow dynamics renormalized by mode coupling controls the configurational 
diffusion rate. 

We thank D.\ Thirumalai and E.\ Shakhnovich for making us aware of 
their referenced works prior to publication.
We also thank K.\ S.\ Schweizer for stimulating discussions.
S.T.\ is a Postdoctoral Fellow for Research Abroad of the Japan Society for 
the Promotion of Science. J.J.P.\ is supported by NSF grant CHE 
95-30680.
P.G.W.\ is supported by NIH grant PHS 5 R01 GM44557 and was a 
Scholar-in-Residence at the Forgaty International Center for 
Advanced Study in the Health Sciences, 
National Institues of Health, Bethesda, U.\ S.\ A.\  when this
work was completed.



\par\noindent
\vfill

\noindent
{Figure 1: Edwards-Anderson order parameters $q_p$ for $p = 1, 50, 200,$ and 
$512$ as 
a function of $\mu=(\beta b)^2$. Parameters used are $N=1024$, 
$\Gamma =1$, $\sigma=1$ and $\eta=0.8$.}

\noindent
{Figure 2: $\mu_{\rm A}=(b/k_BT_{\rm A})^2$ and $\mu_c$ as a function of chain length $N$.
$\mu_{\rm A}$ (solid curve) and $\mu_C$ (dashed curve) are actually 
computed at chain length $2^p$ with $p=5,\cdots,15$.
$\eta=0$ for $\Theta$ solvent and $\eta =0.6, 0.8,$ and $1.0$ for collapsed state. 
Parameters used are $\Gamma=\sigma=1$.}

\noindent
{Figure 3: Edwards-Anderson order parameters $q_p$ and $Q_{i-j}$ 
as a function of $p$ and $i-j$ (solid curves) and 
vector components $\nu_p$ and $\nu_{i-j}$ of the 
dangereous mode in $p$ and $i-j$ representations (dashed curves). 
Parameters used are 
the same as Fig.1 and $\mu$ is slightly above $\mu_{\rm A}$.}

\parskip 20pt \par\noindent

\begin{thebibliography}{99}
\bibitem{Bryngelson95}
Bryngelson, J. D., Onuchic, J. N., Socci, N. D., and Wolynes, P. G., (1995) 
{\it PROTEINS: Struct. Funct. Genetics.} {\bf 21}, 167-195. 
\bibitem{Kinzelbach}
For random directed polymers see Kinzelbach, H. and Horner, H., (1993) 
{\it J. Phys. I France} 
{\bf 3}, 1329-1357.
\bibitem{Timoshenko}
Timoshenko, E. G., Kuznetsov, Yu. A., and Dawson, K. A., (1996) 
{\it Phys. Rev. E.} {\bf 54}, 4071-4086.
\bibitem{Roan}
Roan, J. -R. and Shakhnovich, E. I., (1996) {\it Phys. Rev. E.}
 {\bf 54}, 5340-5357.
\bibitem{Thirumalai}
Thirumalai, D., Ashwin, V., and Bhattacharjee, J. K., (1996) 
to appear {\it Phys. Rev. Lett.}
\bibitem{Takada96}
Takada, S. and Wolynes, P. G., submitted.
\bibitem{Takada97}
Takada, S. and Wolynes, P. G., submitted.
\bibitem{Kirkpatrick87}
Kirkpatrick, T. R. and Wolynes, P. G., (1987) {\it Phys. Rev. B.}
 {\bf  36}, 8552-8564.
\bibitem{Kirkpatrick} 
a)Kirkpatrick, T. R. and Thirumalai, D., (1987) {\it Phys. Rev. B.} 
{\bf  36}, 5388-5397 ;
b)Crisanti, A., Horner, H., and Sommers, H.-J., (1993) 
{\it Z. Phys. B.} {\bf 92}, 257-271. 
\bibitem{Plotkin}
Wang, J., Plotkin, S., and Wolynes, P. G., (1997) {\it J. Phys. I France} 
to appear.
\bibitem{Fischer}
Fischer, K. H. and Hertz, J. A., (1991) {\sl Spin Glasses}, (Cambridge Univ. Press, UK).
\bibitem{Kawasaki}
Kawasaki, K., (1976) in {\it Phase Transitions and Critical Phenomena} 
Vol. 5a, edited by Domb, C. and Green, M. E., (Academic Press, N.Y.), 
pp. 165-403.
\bibitem{Schweizer}
Schweizer, K. S., (1989) {\it J. Chem. Phys.} {\bf 91}, 5802-5821 ; 
 {\sl ibid}, 5822-5839 ; (1993) {\it Physica Scripta} {\bf T49}, 
99-106.
\bibitem{Garel}
Garel, T., Leibler, L., and Orland, H., (1994) {\it J. Phys. II France} 
{\bf 4}, 2139-2148.
\bibitem{onB}
In globule states, $R_g^2=3/4 \sqrt{2/B}$\cite{Sasai90}. $R_g$ is chosen 
by the relation $\eta 4\pi R_g^3/3=N\sigma^3$, where $\eta$ is 
the packing fraction.
The explicit consideration of three body terms will be presented elsewhere\cite{Portman}.
\bibitem{Sasai90}
Sasai, M. and Wolynes, P. G., (1990) {\it Phys. Rev. Lett.} 
{\bf 65}, 2740-2743 ; (1992) {\it Phys. Rev. A.} 
{\bf  46}, 7979-7997.
\bibitem{Portman}
Portman, J. J., Takada, S., and Wolynes, P. G., in preparation.
\bibitem{Gotze}
G\"{o}tze, W., (1991) 
in {\sl Liquids, Freezing and the Glass transition}, edited by  Levesque, D., 
Hansen, J. P., and Zinn-Justin, J., (Elsevier, New York), pp. 287-504.  
\end{thebibliography}
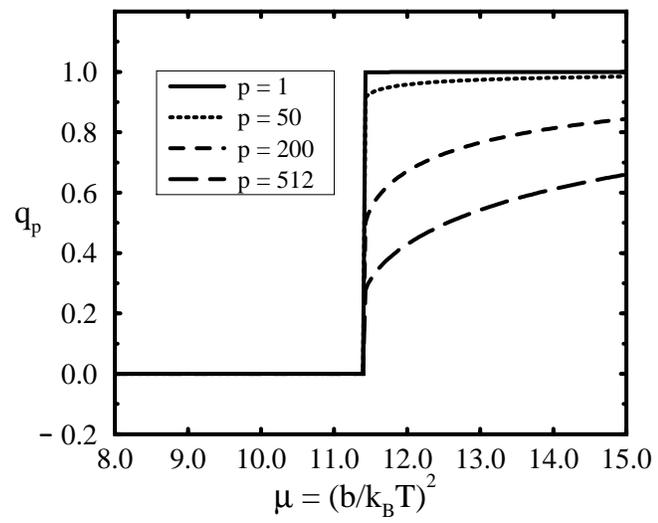
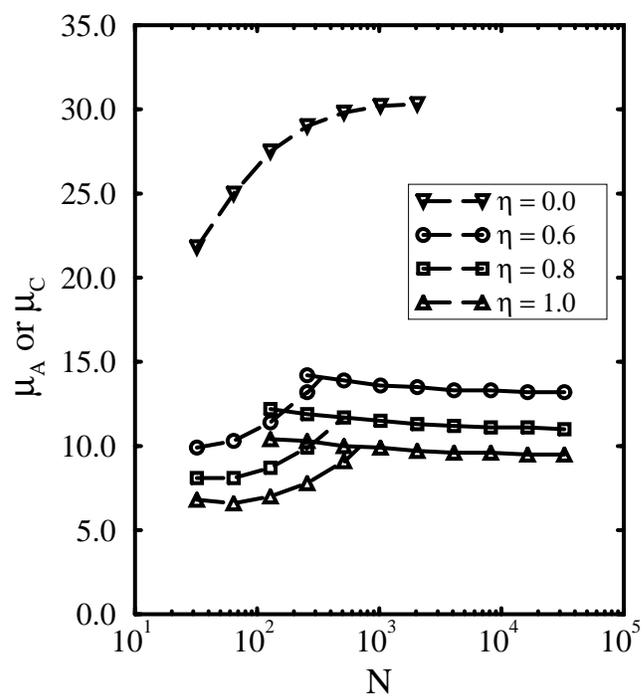
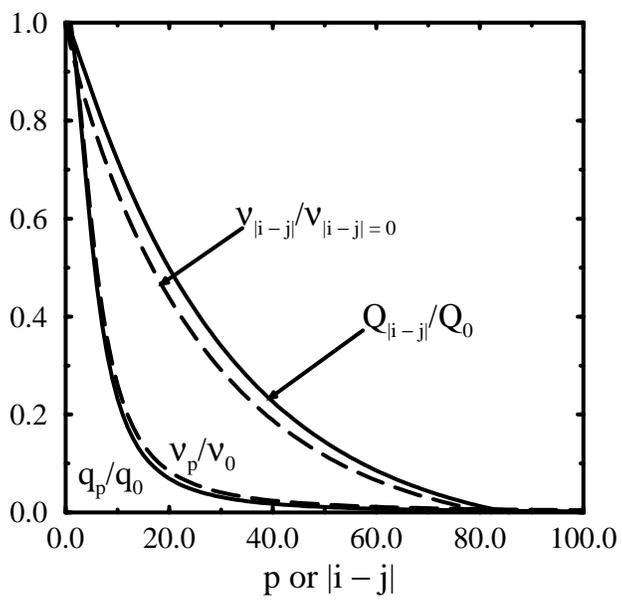
\end{document}